\documentclass[prl,twocolumn,floatfix,superscriptaddress]{revtex4-1}
\usepackage{dcolumn,amsmath}
\usepackage{graphicx}
\usepackage{bm}
\usepackage{hyperref}

\begin{document}
\title{HfF$^+$ as a candidate to search for the nuclear weak quadruple moment}

\author{L.V.\ Skripnikov}\email{leonidos239@gmail.com}
\author{A.N.\ Petrov}
\author{A.V.\ Titov}
\homepage{http://www.qchem.pnpi.spb.ru}
\affiliation{National Research Centre ``Kurchatov Institute'' B.P. Konstantinov Petersburg Nuclear Physics Institute, Gatchina, Leningrad District 188300, Russia}
\affiliation{Saint Petersburg State University, 7/9 Universitetskaya nab., St. Petersburg, 199034 Russia}
\author{V. V. Flambaum}
\affiliation{School of Physics, The University of New South Wales, Sydney NSW 2052, Australia}
\affiliation{Johannes Gutenberg-Universit\"at Mainz, 55099 Mainz, Germany}

\date{19.09.2018}

\begin{abstract}
Nuclei with a quadrupole deformation such as  $^{177}$Hf  have enhanced weak quadrupole moment which induces the tensor weak electron-nucleus interaction in atoms and molecules. Corresponding parity non-conserving (PNC)  effect  is strongly enhanced in the $^3\Delta_1$ electronic state of the $^{177}$HfF$^+$~cation which has very close opposite parity levels mixed by this tensor interaction. In the present paper we perform relativistic many-body calculations of this PNC effect. It is shown that the tensor weak interaction induced by the weak quadrupole moment gives the dominating contribution to the PNC effects in  $^{177}$HfF$^+$ which significantly exceeds contributions of  the  vector anapole moment and the scalar weak charge.
The anapole and the weak charge can contribute due to the nonadiabatic mechanism proposed here.
 Therefore, corresponding experiment will allow one to separate the tensor weak PNC effect from the other PNC effects and to measure  the quadrupole moment of the neutron distribution which gives the dominating contribution to the weak quadrupole moment. 
   \end{abstract}

\maketitle

\section{Introduction}

Parity nonconserving (PNC) effects are a way to test the Standard Model as well as new physics outside it (\cite{Safronova:18,Ginges:04}). They have been extensively studied theoretically and experimentally in atoms (see e.g. Refs.\cite{Safronova:18,Ginges:04,Chubukov:18}). Considerable findings were devoted to search for the weak scalar charge and vector anapole moment of nuclei. 

In Ref.~\cite{Flambaum:17} the effect of the \textit{tensor} contribution to the PNC electron-nucleus interaction was calculated in atoms. It was shown that corresponding experiments will allow one to investigate the quadrupole moment $Q_{n}$ of neutrons in nuclei.
Note that the nuclear \textit{electric} quadrupole moment is induced by nonspherical distribution of charged particles and can be measured by investigating hyperfine splittings in atomic systems. However, no such methods are available for $Q_{n}$ measurement. Previously, there were some studies devoted to investigation of the spherical distribution of neutrons compared to protons, the neutron skin effect \cite{Abrahamyan:2012,Clark:2003}.

It was shown in Refs. \cite{Sushkov:78,Labzowsky:78} that PNC effects are enhanced within diatomic molecules as they have closely spaced rotational levels of opposite parity (e.g. in the HgH$^+$ molecule \cite{Geddes:18}). Recently the experimental progress in investigating systematic effect has been achieved on the $^{180}$Ba$^{19}$F molecule \cite{DeMille:2018}  where the vector weak interactions were considered. In the experiment the external magnetic field has been applied to the molecules to Zeeman shift opposite parity levels up to almost degeneracy. Upper limit on the anapole moment of the $^{19}$F atom has been obtained.

A significant progress has been also achieved in the measurement of effects of fundamental time-reversal and spatial parity symmetry violation (T,P-odd effects) using the $^{180}$HfF$^+$ cation \cite{Cornell:2017} ($^{180}$Hf has zero nucleus spin). The experimental data \cite{Cornell:2017} combined with theoretical data \cite{Petrov:07a,Skripnikov:17a, Fleig:17, Petrov:18} 
can be used to set the upper bound on the electron electric dipole moments and the constant of the scalar-speudoscalar nuclear-electron interaction.
It was shown in Refs. \cite{FDK14,Skripnikov:17b,Petrov:18a} that the $^{177}$HfF$^+$  cation with the $^{177}$Hf isotope having spin $I=3.5$ can be used to search for the T,P-odd nuclear magnetic quadrupole moment. 

In the present paper we calculate a tensor weak interaction effect in the ground rotational level of the first excited metastable electronic  state $^3\Delta_1$ of the $^{177}$HfF$^+$~cation. Attractive feature of the HfF$^+$ cation (which was exploited in the T,P-odd experiment \cite{Cornell:2017}) is that there are two opposite parity states with very small energy interval ($\sim 10$~MHz, see below) leading to the enhancement of the effect. This can be used also to search for the PNC effect induced by the tensor weak interaction \cite{Flambaum:2016}. As far as we know the PNC tensor weak interaction effect has never been calculated or even estimated in molecules.

\section{Theory}

PNC Hamiltonian is given by \cite{Khriplovich:91}:
\begin{equation}
\label{Hpnc}
h_{PNC} = -\frac{G_F}{2\sqrt{2}}\gamma_{5}\left[Zq_{w, p}\rho_{p}({\bf r}) + Nq_{w, n}\rho_{n}({\bf r})\right],
\end{equation}
where Z is the nuclear charge, N in the neutron number, $\gamma_5$ is the Dirac matrix, G$_F\approx2.2225\times 10^{-14}$ a.u. is the Fermi constant, $\rho_{p}(r)$ and $\rho_{n}(r)$ are the density distribution of protons and neutrons normalized to unity; $q_{w, p}$ and $q_{w, n}$ are the weak charges of the proton and neutron, respectively:
\begin{equation}
q_{w, p}\approx 1 - 4\sin^2\theta_W \approx 0.08, \\
\end{equation}
\begin{equation}
q_{w,n} = -1.
\end{equation}
Following Ref.~\cite{Flambaum:17} one can assume $\rho_{p}({\bf r}) \approx \rho_{0p}(r) + \rho_{2p}, (r)Y_{20}(\theta,\phi)$, $\rho_{n}({\bf r}) \approx \rho_{0n}(r) + \rho_{2n}(r)Y_{20}(\theta,\phi)$
(it is taken into account that if the nuclear spin has fixed projection on the z axis the quadrupole part of the density is proportional to $Y_{20}$).
Assuming also $\rho_{0p}(r)=\rho_{0n}(r)=\rho_{0}(r)$ and proportionality between $\rho_{2n}(r)$ and $\rho_{0}(r)$ one obtains the following expression for the tensor part of the PNC interaction~\cite{Flambaum:17}:
\begin{equation}
\label{HQ}
h_{\rm Q}=-\frac{5G_F}{2\sqrt{2}\langle r^2 \rangle}\Sigma_q(-1)^qT^{(2)}_qQ^{TW}_{-q}, 
\end{equation}
where $T^{(2)}_q=C^{(2)}_q\gamma_5 \rho_0(r)$ is the electronic part of the operator,
$C^{(2)}_q=\sqrt{4\pi/5}Y_{2q}$, $\langle r^2\rangle = 4 \pi \int\rho_{0} r^4~dr \approx 3R_N^2/5$ is the mean squared nuclear radius, $R_N$ is the nuclear radius. Here the weak quadrupole moment $Q^{TW} =q_{w,n}Q_n +  q_{w,p}Q_p=-Q_n+0.08 Q_p$ is introduced.

For the case of the considered $^3\Delta_1$ state we introduce the molecular constant
\begin{equation}
\label{WQ}
W_q=\langle ^3\Delta_{+1} |\frac{5 G_F}{2\sqrt{2}  \langle r^2 \rangle }
C^{(2)}_2 \gamma_5 \rho_0(r) | ^3\Delta_{-1} \rangle.
\end{equation}
This constant is the analog of the $W_a$ constant that is required for the interpretation of molecular experiments in terms of the nuclear anapole moment \cite{Flambaum:85b,Kozlov:95,DeMille:08,Kudashov:14} (see below).

P-odd interaction with the nuclear anapole moment is given by the following Hamiltonian \cite{Flambaum:85b}: 
\begin{equation}
h_{\rm A} =  
(W_{a}~\kappa)~\mathbf{n\times S}^{\prime}\cdot
\mathbf{I},
\label{HEFF}
\end{equation}
where $\mathbf{n}$ is the unit vector directed from the heavy nucleus to the light one, $\kappa$ is the dimensionless constant determined by the nuclear anapole moment,  $\bm S'$ is the effective electron  spin ~\cite{KL95}, $W_{a}$ is the parameter determined by the electronic structure of a molecule. The Hamiltonian (\ref{HEFF}) couples electronic states with $\Delta\Omega=\pm1$ ($\Omega$ is the projection of the total electronic moment on the internuclear 
axis. For example for $\Omega=1/2$ states (e.g. the BaF molecule in the ground electronic state) $W_{a}$ is defined by the following expression
\begin{equation}
     W_{a}=\frac{G_\mathrm{F}}{\sqrt{2}}
     \left\langle \Psi_{\Omega=1/2} \left\vert
     \rho_0(r) {\alpha_+}
     \right\vert \Psi_{\Omega=-1/2} \right\rangle,
\label{W_a}
\end{equation} 
where $\alpha_+=\alpha_x+i\cdot \alpha_y$ is the Dirac matrix.
Electronic matrix element between $\Omega=+1$ and $\Omega=-1$ states (the present case) is zero. However, the vector PNC effect can contribute via the interference with the nonadiabatic effects due to the electron-rotation interaction through the intermediate $\Omega=0$ states.

The PNC interaction induced by the scalar weak charge is given by the following Hamiltonian 
\begin{eqnarray}
\label{HpncScal}
\nonumber
h_{\rm Z} = -\frac{G_F}{2\sqrt{2}}\gamma_{5}\left[Zq_{w, p}\rho_{0p}(r) + Nq_{w, n}\rho_{0n}(r)\right]\approx \\
-Q^W\frac{G_F}{2\sqrt{2}}\gamma_{5}\rho_{0}(r),
\end{eqnarray}
where $Q^W$ is the scalar weak charge of the $^{177}$Hf nucleus. Hamiltonian (\ref{HpncScal}) can couple electronic states only with $\Delta\Omega=0$.  However, due to the non-adiabatic interactions scalar weak charge can contribute to the considered PNC amplitude. Below we consider contributions from PNC effects induced by the tensor weak charge, vector anapole and scalar weak charge.

\section{Electronic structure calculation details}

The goal of electronic structure calculation is to calculate PNC matrix elements.

For this we used the two-step approach \cite{Petrov:02,Titov:06amin,Skripnikov:15b,Skripnikov:16a}.
This method allows one to avoid direct 4-component relativistic treatment and use highly accurate wavefunction inside the nucleus which is required for calculation of the matrix elements (\ref{WQ}) and (\ref{W_a}) containing integration over nucleus density.
At the first stage, one considers the valence and outer-core part of the molecular wave function within the generalized relativistic effective core potential (GRECP) method \cite{Titov:99,Mosyagin:10a,Mosyagin:16}. The inner-core electrons are excluded from the explicit treatment. The feature of this stage is that the valence wave functions (spinors) are smoothed in the spatial inner core region of a considered heavy atom. This leads to considerable computational savings. Additional technical advantage is that one can also use very compact contracted basis sets \cite{Skripnikov:16b,Skripnikov:13a}. 
At the second step, one uses the nonvariational procedure developed in \cite{Petrov:02,Titov:06amin,Skripnikov:15b,Skripnikov:16a,Skripnikov:11a} to restore the correct 4-component  behavior of the valence wave function in the spatial core region of a heavy atom. The procedure is based on a proportionality of the valence and low-lying virtual spinors in the inner-core regions of heavy atoms
and has been recently applied to study a number of diatimics~\cite{Skripnikov:14c,Skripnikov:14a,Kudashov:13,Skripnikov:17c,Skripnikov:15d}.

To treat electron correlation effects we used the multireference linear-response coupled cluster
 method with single and double cluster amplitudes \cite{Kallay:3,Kallay:5,Kallay:2} for calculation of the tensor PNC matrix element (\ref{WQ}) and multireference configuration interaction method to treat vector and scalar PNC matrix elements. In the correlation calculations 20 electrons of Hf and F were included explicitly while 60 ($1s..4f$) electrons of Hf were modeled by the GRECP operator.
 For the calculations we have used the [12,16,16,10,8]/(6,5,5,3,1) basis set for Hf and [14,9,4,3]/(4,3,2,1) ANO-I basis set for F \cite{Roos:05} used previously~\cite{Petrov:07a,Petrov:09b,Skripnikov:08a,Petrov:17b}.
 The uncertainty of the calculation for the considered off-diagonal electronic matrix element (\ref{WQ}) (which has two-electron excitation as a leading contributions and therefore determined by pure correlation effects) can be estimated as 50\% which is enough for the current purposes.

For the Hartree-Fock calculations and integral transformations we used the {\sc dirac15} code \cite{DIRAC15}. Relativistic correlation calculations were performed within the {\sc mrcc} code  \cite{MRCC2013}.

\section{Evaluation of molecular properties}

Hafnium isotope $^{177}$Hf has nuclear spin $I^1=7/2$, whereas Fluorine isotope $^{19}$F has nuclear spin $I^2=1/2$. In this paper we use coupling scheme
\begin{equation}
{\bf F}_1 = {\bf J} + {\bf I}^1
\end{equation}
\begin{equation}
{\bf F} = {\bf F}_1 + {\bf I}^2
\end{equation}
where {\bf J} is the total molecular 
less nuclear spins angular momentum.
The field-free energy levels of the ground rotational state with quantum
number $J=1$ splits by the hyperfine interaction with hafnium nucleus into
three groups which are characterized by $F_1=9/2$, $F_1=7/2$, $F_1=5/2$ quantum numbers. The Hyperfine interaction with fluorine nucleus further splits levels with total momentum $F=F_1\pm1/2$.
Note that $F_1$ is not exact but a good quantum number since the
hyperfine interaction with fluorine nucleus is much weaker than the hyperfine interaction with hafnium ones. Finally each hyperfine level has two parity eigenstates known as the $\Omega$-doublet.
These states are equal mixture of $\Omega=\pm1$ states.

Following Refs. \cite{Petrov:11,Petrov:14} energy levels and matrix elements were calculated on wavefunctions of $^{177}$Hf$^{19}$F$^+$ obtained by numerical diagonalization of the molecular Hamiltonian over the basis set of the electronic-rotational-nuclear spin wavefunctions. Details of the Hamiltonian are given in \cite{Petrov:18a,Petrov:17b}.

\section{Results and discussions}
Calculated value of the molecular constant which characterizes the tensor weak PNC interaction in molecule (given by Eq.~(\ref{WQ})) is $W_q = i\cdot 6.0\times10^{-12}$ a.u. Taking $Q_p$=3365 millibarn for $^{177}$Hf and assuming $Q_n\approx N/Z\cdot Q_p$ \cite{Flambaum:17} one obtains $Q^{TW}\approx -4.91$ barn
which is close to more accurate calculation, $-5.44$ barn, given in Ref.~\cite{Lackenby:2018}.
Thus, finaly we have $W_qQ^{TW}\approx 8\times10^{-3}$ Hz.

Table I gives calculated transition matrix elements of the quadrupole weak interaction, $h_Q$, in terms of 
$W_Q Q^{TW}$ as well as the dipole moment between components of the $\Omega$-doublet and splitting for hyperfine sublevels of the ground rotational level of the $^3\Delta_1$ electronic state of $^{177}$HfF$^+$.

As was noted above, the vector PNC interaction induced by the nuclear anapole moment can also contribute to the considered effects via the interference with the nonadiabatic (Coriolis) interaction. The latter electron-rotation interaction can couple states with $\Delta\Omega=\pm1$. We have estimated the effect by numerical diagonalization of the corresponding spin-rotational Hamiltonian. For this the interaction through the $^3\Pi_{0^+}$ and $^3\Pi_{0^-}$ intermediate states has been considered. 
Corresponding contributions are given in Table I in terms of the following molecular electronic constants:
$$
     W^{(1)}_a=\frac{G_\mathrm{F}}{\sqrt{2}}
     \left\langle ^3\Delta_{1} \left\vert
     \rho_0(r) {\alpha_+}
     \right\vert ^3\Pi_{0^+} \right\rangle \approx i\cdot 21~{\rm Hz},
$$
$$
     W^{(2)}_a=\frac{G_\mathrm{F}}{\sqrt{2}}
     \left\langle ^3\Delta_{1} \left\vert
     \rho_0(r) {\alpha_+}
     \right\vert ^3\Pi_{0^-} \right\rangle \approx i\cdot 17~{\rm Hz}.
$$
Using estimation $\kappa(^{177}$Hf$)\approx 0.1$ \cite{Flambaum:84} one obtains that the tensor weak interaction is 1-2 orders of magnitude larger than the effect induced by the nuclear anapole moment.

The contribution of the PNC interaction induced by scalar weak charge has also been considered.
As was noted above this interaction can couple electronic molecular states with $\Delta\Omega=0$.
In our case $^3\Pi_{0^+}$ and $^3\Pi_{0^-}$ electronic states can be mixed by this interaction.
Therefore, this interaction can contribute to the mixing of $^3\Delta_1$ and $^3\Delta_{-1}$ states via the interference with nonadiabatic effects (in the second order of perturbation theory with respect to the latter interaction). Table I gives resulting contribution of the effect due to the scalar weak interaction
 in terms of the following molecular electronic constant:
$$
     W_z=
     \left\langle ^3\Pi_{0^+} \left\vert
      {h_z}
     \right\vert ^3\Pi_{0^-} \right\rangle \approx i\cdot 5~{\rm Hz}.
$$ 
Taking $Q_{W} \approx -105$ one obtains that the resulting contribution is about 2 orders of magnitude smaller than the tensor weak interaction but of the same order as the nuclear anapole moment induced interaction.


\begin{table*}[!h]
\caption{
Transition dipole moment (in a.u. for m$_F$=1$^a$), off-diagonal matrix elements of $h_Q$ (in units $W_qQ^{TW}$), $h_Z$ (in units $10^{-8}W_z Q^W$) between components of $\Omega$-doublets and $\Omega$-doublet splittings, $\Delta$ (in MHz), for the ground rotational $J = 1$ hyperfine levels of the $^3\Delta_1$ electronic state of $^{177}$Hf$^{19}$F$^+$. Individual contributions from the Coriolis interaction with $^3\Pi_{0^+}$ (in units $10^{-3}W_a^{(1)}\kappa$) and $^3\Pi_{0^-}$
(in units $10^{-3}W_a^{(2)}\kappa$) are given for $h_a$.
}
\label{TResult2}
\begin{tabular}{ r  d  d  d  d  d  d}
\hline
	$F_1$ $F$ & d   &        h_Q   & h_a(^3\Pi_{0^+})  &  h_a(^3\Pi_{0^-})  &   h_Z    &   \Delta          \\
\hline
 9/2  5   &    -0.1660  &     -0.0960  &    0.1280           &    -0.1506       & -0.8405   &        3.6             \\
      4   &    -0.2030  &     -0.0961  &    0.1282           &    -0.1508       & -0.8407   &        3.6             \\
 7/2  4   &     0.0618  &      0.3335  &   -0.0613           &     0.0721       & -0.8427   &       -15.7             \\
      3   &     0.0803  &      0.3337  &   -0.0611           &     0.0719       & -0.8430   &       -15.7            \\
 5/2  2   &     0.2310  &     -0.2843  &   -0.2009           &     0.2364       & -0.8509   &       12.0             \\
      3   &     0.1656  &     -0.2847  &   -0.2010           &     0.2366       & -0.8506   &       12.1             \\
\hline
\end{tabular}
\\
$^a$  Values for transition dipole moment for other m$_F$ can be obtained by multiplying on m$_F$,  matrix elements of $h_Q$, $h_A$, $h_Z$ between components of $\Omega$-doublets and $\Omega$-doublet splittings are independent of m$_F$ quantum number.
\end{table*}

\section{Conclusion}
In the present paper the effect of the tensor weak interaction induced by the weak nuclear quadrupole moment
has been calculated in the $^3\Delta_1$ electronic state of the $^{177}$HfF$^+$ cation. 
Weak nuclear quadrupole moment is mainly determined by neutrons quadrupole 
distribution which is unknown and is of interest for the nuclear structure theory. 
It is shown that the tensor weak interaction gives the largest contribution with 
respect to other PNC effects induced by the nuclear anapole moment and nuclear weak charge. 
Thus, it is expected that in the corresponding experiment it will be possible to separate 
the tensor weak PNC effect from the other PNC effects.

\section{Acknowledgement}
Electronic structure calculations were performed at the PIK data center of NRC ``Kurchatov Institute'' -- PNPI. Electronic structure calculations were supported by 
BASIS foundation, according to research Project No.~18-1-3-55-1.  
Molecular calculations are supported by the Russian Science Foundation grant No. 18-12-00227. The calculations of the nuclear structure are supported by Australian Research Council and New Zealand Institute for Advanced Study.


\begin{thebibliography}{53}
\expandafter\ifx\csname natexlab\endcsname\relax\def\natexlab#1{#1}\fi
\expandafter\ifx\csname bibnamefont\endcsname\relax
  \def\bibnamefont#1{#1}\fi
\expandafter\ifx\csname bibfnamefont\endcsname\relax
  \def\bibfnamefont#1{#1}\fi
\expandafter\ifx\csname citenamefont\endcsname\relax
  \def\citenamefont#1{#1}\fi
\expandafter\ifx\csname url\endcsname\relax
  \def\url#1{\texttt{#1}}\fi
\expandafter\ifx\csname urlprefix\endcsname\relax\def\urlprefix{URL }\fi
\providecommand{\bibinfo}[2]{#2}
\providecommand{\eprint}[2][]{\url{#2}}

\bibitem[{\citenamefont{Safronova et~al.}(2018)\citenamefont{Safronova, Budker,
  DeMille, Kimball, Derevianko, and Clark}}]{Safronova:18}
\bibinfo{author}{\bibfnamefont{M.~S.} \bibnamefont{Safronova}},
  \bibinfo{author}{\bibfnamefont{D.}~\bibnamefont{Budker}},
  \bibinfo{author}{\bibfnamefont{D.}~\bibnamefont{DeMille}},
  \bibinfo{author}{\bibfnamefont{D.~F.~J.} \bibnamefont{Kimball}},
  \bibinfo{author}{\bibfnamefont{A.}~\bibnamefont{Derevianko}},
  \bibnamefont{and} \bibinfo{author}{\bibfnamefont{C.~W.} \bibnamefont{Clark}},
  \bibinfo{journal}{Rev. Mod. Phys.} \textbf{\bibinfo{volume}{90}},
  \bibinfo{pages}{025008} (\bibinfo{year}{2018}),
  \urlprefix\url{https://link.aps.org/doi/10.1103/RevModPhys.90.025008}.

\bibitem[{\citenamefont{Ginges and Flambaum}(2004)}]{Ginges:04}
\bibinfo{author}{\bibfnamefont{J.~S.~M.} \bibnamefont{Ginges}}
  \bibnamefont{and} \bibinfo{author}{\bibfnamefont{V.~V.}
  \bibnamefont{Flambaum}}, \bibinfo{journal}{Phys.\ Rep.}
  \textbf{\bibinfo{volume}{397}}, \bibinfo{pages}{63} (\bibinfo{year}{2004}).

\bibitem[{\citenamefont{Chubukov et~al.}(2018)\citenamefont{Chubukov,
  Skripnikov, and Labzowsky}}]{Chubukov:18}
\bibinfo{author}{\bibfnamefont{D.~V.} \bibnamefont{Chubukov}},
  \bibinfo{author}{\bibfnamefont{L.~V.} \bibnamefont{Skripnikov}},
  \bibnamefont{and} \bibinfo{author}{\bibfnamefont{L.~N.}
  \bibnamefont{Labzowsky}}, \bibinfo{journal}{Phys. Rev. A}
  \textbf{\bibinfo{volume}{97}}, \bibinfo{pages}{062512}
  (\bibinfo{year}{2018}).

\bibitem[{\citenamefont{Flambaum et~al.}(2017)\citenamefont{Flambaum, Dzuba,
  and Harabati}}]{Flambaum:17}
\bibinfo{author}{\bibfnamefont{V.~V.} \bibnamefont{Flambaum}},
  \bibinfo{author}{\bibfnamefont{V.~A.} \bibnamefont{Dzuba}}, \bibnamefont{and}
  \bibinfo{author}{\bibfnamefont{C.}~\bibnamefont{Harabati}},
  \bibinfo{journal}{Phys. Rev. A} \textbf{\bibinfo{volume}{96}},
  \bibinfo{pages}{012516} (\bibinfo{year}{2017}).

\bibitem[{\citenamefont{Abrahamyan et~al.}(2012)\citenamefont{Abrahamyan,
  Ahmed, Albataineh, Aniol, Armstrong, Armstrong, Averett, Babineau, Barbieri,
  Bellini et~al.}}]{Abrahamyan:2012}
\bibinfo{author}{\bibfnamefont{S.}~\bibnamefont{Abrahamyan}},
  \bibinfo{author}{\bibfnamefont{Z.}~\bibnamefont{Ahmed}},
  \bibinfo{author}{\bibfnamefont{H.}~\bibnamefont{Albataineh}},
  \bibinfo{author}{\bibfnamefont{K.}~\bibnamefont{Aniol}},
  \bibinfo{author}{\bibfnamefont{D.~S.} \bibnamefont{Armstrong}},
  \bibinfo{author}{\bibfnamefont{W.}~\bibnamefont{Armstrong}},
  \bibinfo{author}{\bibfnamefont{T.}~\bibnamefont{Averett}},
  \bibinfo{author}{\bibfnamefont{B.}~\bibnamefont{Babineau}},
  \bibinfo{author}{\bibfnamefont{A.}~\bibnamefont{Barbieri}},
  \bibinfo{author}{\bibfnamefont{V.}~\bibnamefont{Bellini}},
  \bibnamefont{et~al.} (\bibinfo{collaboration}{PREX Collaboration}),
  \bibinfo{journal}{Phys. Rev. Lett.} \textbf{\bibinfo{volume}{108}},
  \bibinfo{pages}{112502} (\bibinfo{year}{2012}).

\bibitem[{\citenamefont{Clark et~al.}(2003)\citenamefont{Clark, Kerr, and
  Hama}}]{Clark:2003}
\bibinfo{author}{\bibfnamefont{B.~C.} \bibnamefont{Clark}},
  \bibinfo{author}{\bibfnamefont{L.~J.} \bibnamefont{Kerr}}, \bibnamefont{and}
  \bibinfo{author}{\bibfnamefont{S.}~\bibnamefont{Hama}},
  \bibinfo{journal}{Phys. Rev. C} \textbf{\bibinfo{volume}{67}},
  \bibinfo{pages}{054605} (\bibinfo{year}{2003}).

\bibitem[{\citenamefont{Sushkov and Flambaum}(1978)}]{Sushkov:78}
\bibinfo{author}{\bibfnamefont{O.~P.} \bibnamefont{Sushkov}} \bibnamefont{and}
  \bibinfo{author}{\bibfnamefont{V.~V.} \bibnamefont{Flambaum}},
  \bibinfo{journal}{Sov.\ Phys.\ --\ JETP} \textbf{\bibinfo{volume}{48}},
  \bibinfo{pages}{608} (\bibinfo{year}{1978}).

\bibitem[{\citenamefont{Labzowsky}(1978)}]{Labzowsky:78}
\bibinfo{author}{\bibfnamefont{L.~N.} \bibnamefont{Labzowsky}},
  \bibinfo{journal}{Sov.\ Phys.\ --\ JETP} \textbf{\bibinfo{volume}{48}},
  \bibinfo{pages}{434} (\bibinfo{year}{1978}).

\bibitem[{\citenamefont{Geddes et~al.}(2018)\citenamefont{Geddes, Skripnikov,
  Borschevsky, Berengut, Flambaum, and Rakitzis}}]{Geddes:18}
\bibinfo{author}{\bibfnamefont{A.~J.} \bibnamefont{Geddes}},
  \bibinfo{author}{\bibfnamefont{L.~V.} \bibnamefont{Skripnikov}},
  \bibinfo{author}{\bibfnamefont{A.}~\bibnamefont{Borschevsky}},
  \bibinfo{author}{\bibfnamefont{J.~C.} \bibnamefont{Berengut}},
  \bibinfo{author}{\bibfnamefont{V.~V.} \bibnamefont{Flambaum}},
  \bibnamefont{and} \bibinfo{author}{\bibfnamefont{T.~P.}
  \bibnamefont{Rakitzis}}, \bibinfo{journal}{Phys. Rev. A}
  \textbf{\bibinfo{volume}{98}}, \bibinfo{pages}{022508}
  (\bibinfo{year}{2018}).

\bibitem[{\citenamefont{Altunta\ifmmode~\mbox{\c{s}}\else \c{s}\fi{}
  et~al.}(2018)\citenamefont{Altunta\ifmmode~\mbox{\c{s}}\else \c{s}\fi{},
  Ammon, Cahn, and DeMille}}]{DeMille:2018}
\bibinfo{author}{\bibfnamefont{E.}~\bibnamefont{Altunta\ifmmode~\mbox{\c{s}}\else
  \c{s}\fi{}}}, \bibinfo{author}{\bibfnamefont{J.}~\bibnamefont{Ammon}},
  \bibinfo{author}{\bibfnamefont{S.~B.} \bibnamefont{Cahn}}, \bibnamefont{and}
  \bibinfo{author}{\bibfnamefont{D.}~\bibnamefont{DeMille}},
  \bibinfo{journal}{Phys. Rev. Lett.} \textbf{\bibinfo{volume}{120}},
  \bibinfo{pages}{142501} (\bibinfo{year}{2018}).

\bibitem[{\citenamefont{Cairncross et~al.}(2017)\citenamefont{Cairncross,
  Gresh, Grau, Cossel, Roussy, Ni, Zhou, Ye, and Cornell}}]{Cornell:2017}
\bibinfo{author}{\bibfnamefont{W.~B.} \bibnamefont{Cairncross}},
  \bibinfo{author}{\bibfnamefont{D.~N.} \bibnamefont{Gresh}},
  \bibinfo{author}{\bibfnamefont{M.}~\bibnamefont{Grau}},
  \bibinfo{author}{\bibfnamefont{K.~C.} \bibnamefont{Cossel}},
  \bibinfo{author}{\bibfnamefont{T.~S.} \bibnamefont{Roussy}},
  \bibinfo{author}{\bibfnamefont{Y.}~\bibnamefont{Ni}},
  \bibinfo{author}{\bibfnamefont{Y.}~\bibnamefont{Zhou}},
  \bibinfo{author}{\bibfnamefont{J.}~\bibnamefont{Ye}}, \bibnamefont{and}
  \bibinfo{author}{\bibfnamefont{E.~A.} \bibnamefont{Cornell}},
  \bibinfo{journal}{Phys. Rev. Lett.} \textbf{\bibinfo{volume}{119}},
  \bibinfo{pages}{153001} (\bibinfo{year}{2017}).

\bibitem[{\citenamefont{Petrov et~al.}(2007)\citenamefont{Petrov, Mosyagin,
  Isaev, and Titov}}]{Petrov:07a}
\bibinfo{author}{\bibfnamefont{A.~N.} \bibnamefont{Petrov}},
  \bibinfo{author}{\bibfnamefont{N.~S.} \bibnamefont{Mosyagin}},
  \bibinfo{author}{\bibfnamefont{T.~A.} \bibnamefont{Isaev}}, \bibnamefont{and}
  \bibinfo{author}{\bibfnamefont{A.~V.} \bibnamefont{Titov}},
  \bibinfo{journal}{Phys.\ Rev.\ A} \textbf{\bibinfo{volume}{76}},
  \bibinfo{pages}{030501(R)} (\bibinfo{year}{2007}).

\bibitem[{\citenamefont{Skripnikov
  et~al.}(2017{\natexlab{a}})\citenamefont{Skripnikov, Maison, and
  Mosyagin}}]{Skripnikov:17a}
\bibinfo{author}{\bibfnamefont{L.~V.} \bibnamefont{Skripnikov}},
  \bibinfo{author}{\bibfnamefont{D.~E.} \bibnamefont{Maison}},
  \bibnamefont{and} \bibinfo{author}{\bibfnamefont{N.~S.}
  \bibnamefont{Mosyagin}}, \bibinfo{journal}{Phys.\ Rev.\ A}
  \textbf{\bibinfo{volume}{95}}, \bibinfo{pages}{022507}
  (\bibinfo{year}{2017}{\natexlab{a}}).

\bibitem[{\citenamefont{Fleig}(2017)}]{Fleig:17}
\bibinfo{author}{\bibfnamefont{T.}~\bibnamefont{Fleig}},
  \bibinfo{journal}{Phys. Rev. A} \textbf{\bibinfo{volume}{96}},
  \bibinfo{pages}{040502} (\bibinfo{year}{2017}).

\bibitem[{\citenamefont{Petrov}(2018)}]{Petrov:18}
\bibinfo{author}{\bibfnamefont{A.~N.} \bibnamefont{Petrov}},
  \bibinfo{journal}{Phys. Rev. A} \textbf{\bibinfo{volume}{97}},
  \bibinfo{pages}{052504} (\bibinfo{year}{2018}).

\bibitem[{\citenamefont{Flambaum et~al.}(2014)\citenamefont{Flambaum, DeMille,
  and Kozlov}}]{FDK14}
\bibinfo{author}{\bibfnamefont{V.~V.} \bibnamefont{Flambaum}},
  \bibinfo{author}{\bibfnamefont{D.}~\bibnamefont{DeMille}}, \bibnamefont{and}
  \bibinfo{author}{\bibfnamefont{M.~G.} \bibnamefont{Kozlov}},
  \bibinfo{journal}{Phys.\ Rev.\ Lett.} \textbf{\bibinfo{volume}{113}},
  \bibinfo{pages}{103003} (\bibinfo{year}{2014}).

\bibitem[{\citenamefont{Skripnikov
  et~al.}(2017{\natexlab{b}})\citenamefont{Skripnikov, Titov, and
  Flambaum}}]{Skripnikov:17b}
\bibinfo{author}{\bibfnamefont{L.~V.} \bibnamefont{Skripnikov}},
  \bibinfo{author}{\bibfnamefont{A.~V.} \bibnamefont{Titov}}, \bibnamefont{and}
  \bibinfo{author}{\bibfnamefont{V.~V.} \bibnamefont{Flambaum}},
  \bibinfo{journal}{Phys.\ Rev.\ A} \textbf{\bibinfo{volume}{95}},
  \bibinfo{pages}{022512} (\bibinfo{year}{2017}{\natexlab{b}}).

\bibitem[{\citenamefont{Petrov et~al.}(2018)\citenamefont{Petrov, Skripnikov,
  Titov, and Flambaum}}]{Petrov:18a}
\bibinfo{author}{\bibfnamefont{A.~N.} \bibnamefont{Petrov}},
  \bibinfo{author}{\bibfnamefont{L.~V.} \bibnamefont{Skripnikov}},
  \bibinfo{author}{\bibfnamefont{A.~V.} \bibnamefont{Titov}}, \bibnamefont{and}
  \bibinfo{author}{\bibfnamefont{V.~V.} \bibnamefont{Flambaum}},
  \bibinfo{journal}{Phys. Rev. A} \textbf{\bibinfo{volume}{98}},
  \bibinfo{pages}{042502} (\bibinfo{year}{2018}).

\bibitem[{\citenamefont{Flambaum}(2016)}]{Flambaum:2016}
\bibinfo{author}{\bibfnamefont{V.~V.} \bibnamefont{Flambaum}},
  \bibinfo{journal}{Phys. Rev. Lett.} \textbf{\bibinfo{volume}{117}},
  \bibinfo{pages}{072501} (\bibinfo{year}{2016}).

\bibitem[{\citenamefont{Khriplovich}(1991)}]{Khriplovich:91}
\bibinfo{author}{\bibfnamefont{I.~B.} \bibnamefont{Khriplovich}},
  \emph{\bibinfo{title}{Parity non-conservation in atomic phenomena}}
  (\bibinfo{publisher}{Gordon and Breach}, \bibinfo{address}{New York},
  \bibinfo{year}{1991}).

\bibitem[{\citenamefont{Flambaum and Khriplovich}(1985)}]{Flambaum:85b}
\bibinfo{author}{\bibfnamefont{V.~V.} \bibnamefont{Flambaum}} \bibnamefont{and}
  \bibinfo{author}{\bibfnamefont{I.~B.} \bibnamefont{Khriplovich}},
  \bibinfo{journal}{Phys.\ Lett.\ A} \textbf{\bibinfo{volume}{110}},
  \bibinfo{pages}{121} (\bibinfo{year}{1985}).

\bibitem[{\citenamefont{Kozlov and Labzowsky}(1995{\natexlab{a}})}]{Kozlov:95}
\bibinfo{author}{\bibfnamefont{M.}~\bibnamefont{Kozlov}} \bibnamefont{and}
  \bibinfo{author}{\bibfnamefont{L.}~\bibnamefont{Labzowsky}},
  \bibinfo{journal}{J.\ Phys.\ B} \textbf{\bibinfo{volume}{28}},
  \bibinfo{pages}{1933} (\bibinfo{year}{1995}{\natexlab{a}}).

\bibitem[{\citenamefont{DeMille et~al.}(2008)\citenamefont{DeMille, Cahn,
  Murphree, Rahmlow, and Kozlov}}]{DeMille:08}
\bibinfo{author}{\bibfnamefont{D.}~\bibnamefont{DeMille}},
  \bibinfo{author}{\bibfnamefont{S.~B.} \bibnamefont{Cahn}},
  \bibinfo{author}{\bibfnamefont{D.}~\bibnamefont{Murphree}},
  \bibinfo{author}{\bibfnamefont{D.~A.} \bibnamefont{Rahmlow}},
  \bibnamefont{and} \bibinfo{author}{\bibfnamefont{M.~G.}
  \bibnamefont{Kozlov}}, \bibinfo{journal}{Phys.\ Rev.\ Lett.}
  \textbf{\bibinfo{volume}{100}}, \bibinfo{pages}{023003}
  (\bibinfo{year}{2008}).

\bibitem[{\citenamefont{Kudashov et~al.}(2014)\citenamefont{Kudashov, Petrov,
  Skripnikov, Mosyagin, Isaev, Berger, and Titov}}]{Kudashov:14}
\bibinfo{author}{\bibfnamefont{A.~D.} \bibnamefont{Kudashov}},
  \bibinfo{author}{\bibfnamefont{A.~N.} \bibnamefont{Petrov}},
  \bibinfo{author}{\bibfnamefont{L.~V.} \bibnamefont{Skripnikov}},
  \bibinfo{author}{\bibfnamefont{N.~S.} \bibnamefont{Mosyagin}},
  \bibinfo{author}{\bibfnamefont{T.~A.} \bibnamefont{Isaev}},
  \bibinfo{author}{\bibfnamefont{R.}~\bibnamefont{Berger}}, \bibnamefont{and}
  \bibinfo{author}{\bibfnamefont{A.~V.} \bibnamefont{Titov}},
  \bibinfo{journal}{Phys.\ Rev.\ A} \textbf{\bibinfo{volume}{90}},
  \bibinfo{pages}{052513} (\bibinfo{year}{2014}).

\bibitem[{\citenamefont{Kozlov and Labzowsky}(1995{\natexlab{b}})}]{KL95}
\bibinfo{author}{\bibfnamefont{M.}~\bibnamefont{Kozlov}} \bibnamefont{and}
  \bibinfo{author}{\bibfnamefont{L.}~\bibnamefont{Labzowsky}},
  \bibinfo{journal}{J.\ Phys.\ B} \textbf{\bibinfo{volume}{28}},
  \bibinfo{pages}{1933} (\bibinfo{year}{1995}{\natexlab{b}}).

\bibitem[{\citenamefont{Petrov et~al.}(2002)\citenamefont{Petrov, Mosyagin,
  Isaev, Titov, Ezhov, Eliav, and Kaldor}}]{Petrov:02}
\bibinfo{author}{\bibfnamefont{A.~N.} \bibnamefont{Petrov}},
  \bibinfo{author}{\bibfnamefont{N.~S.} \bibnamefont{Mosyagin}},
  \bibinfo{author}{\bibfnamefont{T.~A.} \bibnamefont{Isaev}},
  \bibinfo{author}{\bibfnamefont{A.~V.} \bibnamefont{Titov}},
  \bibinfo{author}{\bibfnamefont{V.~F.} \bibnamefont{Ezhov}},
  \bibinfo{author}{\bibfnamefont{E.}~\bibnamefont{Eliav}}, \bibnamefont{and}
  \bibinfo{author}{\bibfnamefont{U.}~\bibnamefont{Kaldor}},
  \bibinfo{journal}{Phys.\ Rev.\ Lett.} \textbf{\bibinfo{volume}{88}},
  \bibinfo{pages}{073001} (\bibinfo{year}{2002}).

\bibitem[{\citenamefont{Titov et~al.}(2006)\citenamefont{Titov, Mosyagin,
  Petrov, Isaev, and DeMille}}]{Titov:06amin}
\bibinfo{author}{\bibfnamefont{A.~V.} \bibnamefont{Titov}},
  \bibinfo{author}{\bibfnamefont{N.~S.} \bibnamefont{Mosyagin}},
  \bibinfo{author}{\bibfnamefont{A.~N.} \bibnamefont{Petrov}},
  \bibinfo{author}{\bibfnamefont{T.~A.} \bibnamefont{Isaev}}, \bibnamefont{and}
  \bibinfo{author}{\bibfnamefont{D.~P.} \bibnamefont{DeMille}},
  \bibinfo{journal}{Progr.\ Theor.\ Chem.\ Phys.}
  \textbf{\bibinfo{volume}{15}}, \bibinfo{pages}{253} (\bibinfo{year}{2006}).

\bibitem[{\citenamefont{Skripnikov and Titov}(2015)}]{Skripnikov:15b}
\bibinfo{author}{\bibfnamefont{L.~V.} \bibnamefont{Skripnikov}}
  \bibnamefont{and} \bibinfo{author}{\bibfnamefont{A.~V.} \bibnamefont{Titov}},
  \bibinfo{journal}{Phys. Rev. A} \textbf{\bibinfo{volume}{91}},
  \bibinfo{pages}{042504} (\bibinfo{year}{2015}).

\bibitem[{\citenamefont{Skripnikov and Titov}(2016)}]{Skripnikov:16a}
\bibinfo{author}{\bibfnamefont{L.~V.} \bibnamefont{Skripnikov}}
  \bibnamefont{and} \bibinfo{author}{\bibfnamefont{A.~V.} \bibnamefont{Titov}},
  \bibinfo{journal}{J.\ Chem.\ Phys.} \textbf{\bibinfo{volume}{145}},
  \bibinfo{eid}{054115} (\bibinfo{year}{2016}).

\bibitem[{\citenamefont{Titov and Mosyagin}(1999)}]{Titov:99}
\bibinfo{author}{\bibfnamefont{A.~V.} \bibnamefont{Titov}} \bibnamefont{and}
  \bibinfo{author}{\bibfnamefont{N.~S.} \bibnamefont{Mosyagin}},
  \bibinfo{journal}{Int.\ J.\ Quantum Chem.} \textbf{\bibinfo{volume}{71}},
  \bibinfo{pages}{359} (\bibinfo{year}{1999}).

\bibitem[{\citenamefont{Mosyagin et~al.}(2010)\citenamefont{Mosyagin,
  Zaitsevskii, and Titov}}]{Mosyagin:10a}
\bibinfo{author}{\bibfnamefont{N.~S.} \bibnamefont{Mosyagin}},
  \bibinfo{author}{\bibfnamefont{A.~V.} \bibnamefont{Zaitsevskii}},
  \bibnamefont{and} \bibinfo{author}{\bibfnamefont{A.~V.} \bibnamefont{Titov}},
  \bibinfo{journal}{Review of Atomic and Molecular Physics}
  \textbf{\bibinfo{volume}{1}}, \bibinfo{pages}{63} (\bibinfo{year}{2010}).

\bibitem[{\citenamefont{Mosyagin et~al.}(2016)\citenamefont{Mosyagin,
  Zaitsevskii, Skripnikov, and Titov}}]{Mosyagin:16}
\bibinfo{author}{\bibfnamefont{N.~S.} \bibnamefont{Mosyagin}},
  \bibinfo{author}{\bibfnamefont{A.~V.} \bibnamefont{Zaitsevskii}},
  \bibinfo{author}{\bibfnamefont{L.~V.} \bibnamefont{Skripnikov}},
  \bibnamefont{and} \bibinfo{author}{\bibfnamefont{A.~V.} \bibnamefont{Titov}},
  \bibinfo{journal}{Int.\ J.\ Quantum Chem.} \textbf{\bibinfo{volume}{116}},
  \bibinfo{pages}{301} (\bibinfo{year}{2016}), ISSN \bibinfo{issn}{1097-461X}.

\bibitem[{\citenamefont{Skripnikov}(2016)}]{Skripnikov:16b}
\bibinfo{author}{\bibfnamefont{L.~V.} \bibnamefont{Skripnikov}},
  \bibinfo{journal}{J.\ Chem.\ Phys.} \textbf{\bibinfo{volume}{145}},
  \bibinfo{pages}{214301} (\bibinfo{year}{2016}).

\bibitem[{\citenamefont{Skripnikov et~al.}(2013)\citenamefont{Skripnikov,
  Mosyagin, and Titov}}]{Skripnikov:13a}
\bibinfo{author}{\bibfnamefont{L.~V.} \bibnamefont{Skripnikov}},
  \bibinfo{author}{\bibfnamefont{N.~S.} \bibnamefont{Mosyagin}},
  \bibnamefont{and} \bibinfo{author}{\bibfnamefont{A.~V.} \bibnamefont{Titov}},
  \bibinfo{journal}{Chem.\ Phys.\ Lett.} \textbf{\bibinfo{volume}{555}},
  \bibinfo{pages}{79} (\bibinfo{year}{2013}).

\bibitem[{\citenamefont{Skripnikov et~al.}(2011)\citenamefont{Skripnikov,
  Titov, Petrov, Mosyagin, and Sushkov}}]{Skripnikov:11a}
\bibinfo{author}{\bibfnamefont{L.~V.} \bibnamefont{Skripnikov}},
  \bibinfo{author}{\bibfnamefont{A.~V.} \bibnamefont{Titov}},
  \bibinfo{author}{\bibfnamefont{A.~N.} \bibnamefont{Petrov}},
  \bibinfo{author}{\bibfnamefont{N.~S.} \bibnamefont{Mosyagin}},
  \bibnamefont{and} \bibinfo{author}{\bibfnamefont{O.~P.}
  \bibnamefont{Sushkov}}, \bibinfo{journal}{Phys.\ Rev.\ A}
  \textbf{\bibinfo{volume}{84}}, \bibinfo{pages}{022505}
  (\bibinfo{year}{2011}).

\bibitem[{\citenamefont{Skripnikov
  et~al.}(2014{\natexlab{a}})\citenamefont{Skripnikov, Kudashov, Petrov, and
  Titov}}]{Skripnikov:14c}
\bibinfo{author}{\bibfnamefont{L.~V.} \bibnamefont{Skripnikov}},
  \bibinfo{author}{\bibfnamefont{A.~D.} \bibnamefont{Kudashov}},
  \bibinfo{author}{\bibfnamefont{A.~N.} \bibnamefont{Petrov}},
  \bibnamefont{and} \bibinfo{author}{\bibfnamefont{A.~V.} \bibnamefont{Titov}},
  \bibinfo{journal}{Phys.\ Rev.\ A} \textbf{\bibinfo{volume}{90}},
  \bibinfo{pages}{064501} (\bibinfo{year}{2014}{\natexlab{a}}).

\bibitem[{\citenamefont{Skripnikov
  et~al.}(2014{\natexlab{b}})\citenamefont{Skripnikov, Petrov, Titov, and
  Flambaum}}]{Skripnikov:14a}
\bibinfo{author}{\bibfnamefont{L.~V.} \bibnamefont{Skripnikov}},
  \bibinfo{author}{\bibfnamefont{A.~N.} \bibnamefont{Petrov}},
  \bibinfo{author}{\bibfnamefont{A.~V.} \bibnamefont{Titov}}, \bibnamefont{and}
  \bibinfo{author}{\bibfnamefont{V.~V.} \bibnamefont{Flambaum}},
  \bibinfo{journal}{Phys.\ Rev.\ Lett.} \textbf{\bibinfo{volume}{113}},
  \bibinfo{pages}{263006} (\bibinfo{year}{2014}{\natexlab{b}}).

\bibitem[{\citenamefont{Kudashov et~al.}(2013)\citenamefont{Kudashov, Petrov,
  Skripnikov, Mosyagin, Titov, and Flambaum}}]{Kudashov:13}
\bibinfo{author}{\bibfnamefont{A.~D.} \bibnamefont{Kudashov}},
  \bibinfo{author}{\bibfnamefont{A.~N.} \bibnamefont{Petrov}},
  \bibinfo{author}{\bibfnamefont{L.~V.} \bibnamefont{Skripnikov}},
  \bibinfo{author}{\bibfnamefont{N.~S.} \bibnamefont{Mosyagin}},
  \bibinfo{author}{\bibfnamefont{A.~V.} \bibnamefont{Titov}}, \bibnamefont{and}
  \bibinfo{author}{\bibfnamefont{V.~V.} \bibnamefont{Flambaum}},
  \bibinfo{journal}{Phys.\ Rev.\ A} \textbf{\bibinfo{volume}{87}},
  \bibinfo{pages}{020102(R)} (\bibinfo{year}{2013}).

\bibitem[{\citenamefont{Skripnikov}(2017)}]{Skripnikov:17c}
\bibinfo{author}{\bibfnamefont{L.~V.} \bibnamefont{Skripnikov}},
  \bibinfo{journal}{J.\ Chem.\ Phys.} \textbf{\bibinfo{volume}{147}},
  \bibinfo{pages}{021101} (\bibinfo{year}{2017}).

\bibitem[{\citenamefont{Skripnikov et~al.}(2015)\citenamefont{Skripnikov,
  Petrov, Titov, Mawhorter, Baum, Sears, and Grabow}}]{Skripnikov:15d}
\bibinfo{author}{\bibfnamefont{L.~V.} \bibnamefont{Skripnikov}},
  \bibinfo{author}{\bibfnamefont{A.~N.} \bibnamefont{Petrov}},
  \bibinfo{author}{\bibfnamefont{A.~V.} \bibnamefont{Titov}},
  \bibinfo{author}{\bibfnamefont{R.~J.} \bibnamefont{Mawhorter}},
  \bibinfo{author}{\bibfnamefont{A.~L.} \bibnamefont{Baum}},
  \bibinfo{author}{\bibfnamefont{T.~J.} \bibnamefont{Sears}}, \bibnamefont{and}
  \bibinfo{author}{\bibfnamefont{J.-U.} \bibnamefont{Grabow}},
  \bibinfo{journal}{Phys. Rev. A} \textbf{\bibinfo{volume}{92}},
  \bibinfo{pages}{032508} (\bibinfo{year}{2015}).

\bibitem[{\citenamefont{K\'{a}llay et~al.}(2003)\citenamefont{K\'{a}llay,
  Gauss, and Szalay}}]{Kallay:3}
\bibinfo{author}{\bibfnamefont{M.}~\bibnamefont{K\'{a}llay}},
  \bibinfo{author}{\bibfnamefont{J.}~\bibnamefont{Gauss}}, \bibnamefont{and}
  \bibinfo{author}{\bibfnamefont{P.~G.} \bibnamefont{Szalay}},
  \bibinfo{journal}{J.\ Chem.\ Phys.} \textbf{\bibinfo{volume}{119}},
  \bibinfo{pages}{2991} (\bibinfo{year}{2003}).

\bibitem[{\citenamefont{K\'{a}llay and Gauss}(2004)}]{Kallay:5}
\bibinfo{author}{\bibfnamefont{M.}~\bibnamefont{K\'{a}llay}} \bibnamefont{and}
  \bibinfo{author}{\bibfnamefont{J.}~\bibnamefont{Gauss}},
  \bibinfo{journal}{J.\ Chem.\ Phys.} \textbf{\bibinfo{volume}{121}},
  \bibinfo{pages}{9257} (\bibinfo{year}{2004}).

\bibitem[{\citenamefont{K\'{a}llay et~al.}(2002)\citenamefont{K\'{a}llay,
  Szalay, and Surj\'{a}n}}]{Kallay:2}
\bibinfo{author}{\bibfnamefont{M.}~\bibnamefont{K\'{a}llay}},
  \bibinfo{author}{\bibfnamefont{P.~G.} \bibnamefont{Szalay}},
  \bibnamefont{and} \bibinfo{author}{\bibfnamefont{P.~R.}
  \bibnamefont{Surj\'{a}n}}, \bibinfo{journal}{J.\ Chem.\ Phys.}
  \textbf{\bibinfo{volume}{117}}, \bibinfo{pages}{980} (\bibinfo{year}{2002}).

\bibitem[{\citenamefont{Roos et~al.}(2005)\citenamefont{Roos, Lindh, {\o
  A}~Malmqvist", Veryazov, and Widmark}}]{Roos:05}
\bibinfo{author}{\bibfnamefont{B.~O.} \bibnamefont{Roos}},
  \bibinfo{author}{\bibfnamefont{R.}~\bibnamefont{Lindh}},
  \bibinfo{author}{\bibfnamefont{P.}~\bibnamefont{{\o A}~Malmqvist"}},
  \bibinfo{author}{\bibfnamefont{V.}~\bibnamefont{Veryazov}}, \bibnamefont{and}
  \bibinfo{author}{\bibfnamefont{P.~O.} \bibnamefont{Widmark}},
  \bibinfo{journal}{J.\ Phys.\ Chem.\ A} \textbf{\bibinfo{volume}{108}},
  \bibinfo{pages}{2851} (\bibinfo{year}{2005}).

\bibitem[{\citenamefont{Petrov et~al.}(2009)\citenamefont{Petrov, Mosyagin, and
  Titov}}]{Petrov:09b}
\bibinfo{author}{\bibfnamefont{A.~N.} \bibnamefont{Petrov}},
  \bibinfo{author}{\bibfnamefont{N.~S.} \bibnamefont{Mosyagin}},
  \bibnamefont{and} \bibinfo{author}{\bibfnamefont{A.~V.} \bibnamefont{Titov}},
  \bibinfo{journal}{Phys.\ Rev.\ A} \textbf{\bibinfo{volume}{79}},
  \bibinfo{pages}{012505} (\bibinfo{year}{2009}).

\bibitem[{\citenamefont{Skripnikov et~al.}(2008)\citenamefont{Skripnikov,
  Mosyagin, Petrov, and Titov}}]{Skripnikov:08a}
\bibinfo{author}{\bibfnamefont{L.~V.} \bibnamefont{Skripnikov}},
  \bibinfo{author}{\bibfnamefont{N.~S.} \bibnamefont{Mosyagin}},
  \bibinfo{author}{\bibfnamefont{A.~N.} \bibnamefont{Petrov}},
  \bibnamefont{and} \bibinfo{author}{\bibfnamefont{A.~V.} \bibnamefont{Titov}},
  \bibinfo{journal}{JETP Letters} \textbf{\bibinfo{volume}{88}},
  \bibinfo{pages}{578} (\bibinfo{year}{2008}).

\bibitem[{\citenamefont{Petrov et~al.}(2017)\citenamefont{Petrov, Skripnikov,
  and Titov}}]{Petrov:17b}
\bibinfo{author}{\bibfnamefont{A.~N.} \bibnamefont{Petrov}},
  \bibinfo{author}{\bibfnamefont{L.~V.} \bibnamefont{Skripnikov}},
  \bibnamefont{and} \bibinfo{author}{\bibfnamefont{A.~V.} \bibnamefont{Titov}},
  \bibinfo{journal}{Phys. Rev. A} \textbf{\bibinfo{volume}{96}},
  \bibinfo{pages}{022508} (\bibinfo{year}{2017}).

\bibitem[{DIR()}]{DIRAC15}
\bibinfo{note}{DIRAC, a relativistic ab initio electronic structure program,
  Release DIRAC15 (2015), written by R. Bast, T. Saue, L. Visscher, and H. J.
  Aa. Jensen, with contributions from V. Bakken, K. G. Dyall, S. Dubillard, U.
  Ekstroem, E. Eliav, T. Enevoldsen, E. Fasshauer, T. Fleig, O. Fossgaard, A.
  S. P. Gomes, T. Helgaker, J. Henriksson, M. Ilias, Ch. R. Jacob, S. Knecht,
  S. Komorovsky, O. Kullie, J. K. Laerdahl, C. V. Larsen, Y. S. Lee, H. S.
  Nataraj, M. K. Nayak, P. Norman, G. Olejniczak, J. Olsen, Y. C. Park, J. K.
  Pedersen, M. Pernpointner, R. Di Remigio, K. Ruud, P. Salek, B.
  Schimmelpfennig, J. Sikkema, A. J. Thorvaldsen, J. Thyssen, J. van Stralen,
  S. Villaume, O. Visser, T. Winther, and S. Yamamoto (see
  http://www.diracprogram.org).}

\bibitem[{MRC()}]{MRCC2013}
\bibinfo{note}{{\sc mrcc}, a quantum chemical program suite written by M.
  K\'{a}llay, Z. Rolik, I. Ladj\'{a}nszki, L. Szegedy, B. Lad\'{o}czki, J.
  Csontos, and B. Kornis. See also Z. Rolik and M. K\'{a}llay, J. Chem. Phys.
  135, 104111 (2011), as well as: www.mrcc.hu}.

\bibitem[{\citenamefont{Petrov}(2011)}]{Petrov:11}
\bibinfo{author}{\bibfnamefont{A.~N.} \bibnamefont{Petrov}},
  \bibinfo{journal}{Phys.\ Rev.\ A} \textbf{\bibinfo{volume}{83}},
  \bibinfo{pages}{024502} (\bibinfo{year}{2011}).

\bibitem[{\citenamefont{Petrov et~al.}(2014)\citenamefont{Petrov, Skripnikov,
  Titov, Hutzler, Hess, O'Leary, Spaun, DeMille, Gabrielse, and
  Doyle}}]{Petrov:14}
\bibinfo{author}{\bibfnamefont{A.~N.} \bibnamefont{Petrov}},
  \bibinfo{author}{\bibfnamefont{L.~V.} \bibnamefont{Skripnikov}},
  \bibinfo{author}{\bibfnamefont{A.~V.} \bibnamefont{Titov}},
  \bibinfo{author}{\bibfnamefont{N.~R.} \bibnamefont{Hutzler}},
  \bibinfo{author}{\bibfnamefont{P.~W.} \bibnamefont{Hess}},
  \bibinfo{author}{\bibfnamefont{B.~R.} \bibnamefont{O'Leary}},
  \bibinfo{author}{\bibfnamefont{B.}~\bibnamefont{Spaun}},
  \bibinfo{author}{\bibfnamefont{D.}~\bibnamefont{DeMille}},
  \bibinfo{author}{\bibfnamefont{G.}~\bibnamefont{Gabrielse}},
  \bibnamefont{and} \bibinfo{author}{\bibfnamefont{J.~M.} \bibnamefont{Doyle}},
  \bibinfo{journal}{Phys. Rev. A} \textbf{\bibinfo{volume}{89}},
  \bibinfo{pages}{062505} (\bibinfo{year}{2014}).

\bibitem[{\citenamefont{Lackenby and Flambaum}(2018)}]{Lackenby:2018}
\bibinfo{author}{\bibfnamefont{B.~G.~C.} \bibnamefont{Lackenby}}
  \bibnamefont{and} \bibinfo{author}{\bibfnamefont{V.~V.}
  \bibnamefont{Flambaum}}, \bibinfo{journal}{Journal of Physics G: Nuclear and
  Particle Physics} \textbf{\bibinfo{volume}{45}}, \bibinfo{pages}{075105}
  (\bibinfo{year}{2018}).

\bibitem[{\citenamefont{Flambaum et~al.}(1984)\citenamefont{Flambaum,
  Khriplovich, and Sushkov}}]{Flambaum:84}
\bibinfo{author}{\bibfnamefont{V.}~\bibnamefont{Flambaum}},
  \bibinfo{author}{\bibfnamefont{I.}~\bibnamefont{Khriplovich}},
  \bibnamefont{and} \bibinfo{author}{\bibfnamefont{O.}~\bibnamefont{Sushkov}},
  \bibinfo{journal}{Physics Letters B} \textbf{\bibinfo{volume}{146}},
  \bibinfo{pages}{367 } (\bibinfo{year}{1984}), ISSN \bibinfo{issn}{0370-2693}.

\end{thebibliography}

\end{document}